\documentclass[aps,preprint,superscriptaddress,showpacs,byrevtex]{revtex4}

\usepackage{graphicx}
\usepackage{epsfig}
\usepackage{dcolumn}
\usepackage{bm}
\usepackage{rotating}

\newcommand{\lum}{{\cal L}}

\newcommand{\BR}{{\cal B}}

\newcommand{\piz}{\pi^0}

\newcommand{\ks}{K_S^0}

\newcommand{\EE}{e^+e^-}
\newcommand{\MM}{\mu^+\mu^-}

\newcommand{\pp}{\pi^+\pi^-}

\newcommand{\beq}{\begin{equation}}
\newcommand{\eeq}{\end{equation}}
\newcommand{\beqy}{\begin{eqnarray}}
\newcommand{\eeqy}{\end{eqnarray}}
\newcommand{\bitm}{\begin{itemize}}
\newcommand{\eitm}{\end{itemize}}

\begin{document}


\title{ 
\boldmath Measurement of $e^+ e^- \to \omega \pi^0$, $K^{\ast}(892)\bar{K}$
and $K_2^{\ast}(1430)\bar{K}$ at $\sqrt{s}$ near 10.6 GeV}


\noaffiliation
\affiliation{University of the Basque Country UPV/EHU, 48080 Bilbao}
\affiliation{Beihang University, Beijing 100191}
\affiliation{University of Bonn, 53115 Bonn}
\affiliation{Budker Institute of Nuclear Physics SB RAS and Novosibirsk State University, Novosibirsk 630090}
\affiliation{Faculty of Mathematics and Physics, Charles University, 121 16 Prague}
\affiliation{University of Cincinnati, Cincinnati, Ohio 45221}
\affiliation{Deutsches Elektronen--Synchrotron, 22607 Hamburg}
\affiliation{Department of Physics, Fu Jen Catholic University, Taipei 24205}
\affiliation{Justus-Liebig-Universit\"at Gie\ss{}en, 35392 Gie\ss{}en}
\affiliation{II. Physikalisches Institut, Georg-August-Universit\"at G\"ottingen, 37073 G\"ottingen}
\affiliation{Gyeongsang National University, Chinju 660-701}
\affiliation{Hanyang University, Seoul 133-791}
\affiliation{University of Hawaii, Honolulu, Hawaii 96822}
\affiliation{High Energy Accelerator Research Organization (KEK), Tsukuba 305-0801}
\affiliation{Hiroshima Institute of Technology, Hiroshima 731-5193}
\affiliation{Ikerbasque, 48011 Bilbao}
\affiliation{Indian Institute of Technology Guwahati, Assam 781039}
\affiliation{Indian Institute of Technology Madras, Chennai 600036}
\affiliation{Indiana University, Bloomington, Indiana 47408}
\affiliation{Institute of High Energy Physics, Chinese Academy of Sciences, Beijing 100049}
\affiliation{Institute of High Energy Physics, Vienna 1050}
\affiliation{Institute for High Energy Physics, Protvino 142281}
\affiliation{INFN - Sezione di Torino, 10125 Torino}
\affiliation{Institute for Theoretical and Experimental Physics, Moscow 117218}
\affiliation{J. Stefan Institute, 1000 Ljubljana}
\affiliation{Kanagawa University, Yokohama 221-8686}
\affiliation{Institut f\"ur Experimentelle Kernphysik, Karlsruher Institut f\"ur Technologie, 76131 Karlsruhe}
\affiliation{Korea Institute of Science and Technology Information, Daejeon 305-806}
\affiliation{Korea University, Seoul 136-713}
\affiliation{Kyungpook National University, Daegu 702-701}
\affiliation{\'Ecole Polytechnique F\'ed\'erale de Lausanne (EPFL), Lausanne 1015}
\affiliation{Faculty of Mathematics and Physics, University of Ljubljana, 1000 Ljubljana}
\affiliation{Luther College, Decorah, Iowa 52101}
\affiliation{University of Maribor, 2000 Maribor}
\affiliation{Max-Planck-Institut f\"ur Physik, 80805 M\"unchen}
\affiliation{School of Physics, University of Melbourne, Victoria 3010}
\affiliation{Moscow Physical Engineering Institute, Moscow 115409}
\affiliation{Moscow Institute of Physics and Technology, Moscow Region 141700}
\affiliation{Graduate School of Science, Nagoya University, Nagoya 464-8602}
\affiliation{Kobayashi-Maskawa Institute, Nagoya University, Nagoya 464-8602}
\affiliation{Nara Women's University, Nara 630-8506}
\affiliation{National Central University, Chung-li 32054}
\affiliation{National United University, Miao Li 36003}
\affiliation{Department of Physics, National Taiwan University, Taipei 10617}
\affiliation{H. Niewodniczanski Institute of Nuclear Physics, Krakow 31-342}
\affiliation{Nippon Dental University, Niigata 951-8580}
\affiliation{Niigata University, Niigata 950-2181}
\affiliation{Osaka City University, Osaka 558-8585}
\affiliation{Pacific Northwest National Laboratory, Richland, Washington 99352}
\affiliation{Panjab University, Chandigarh 160014}
\affiliation{University of Pittsburgh, Pittsburgh, Pennsylvania 15260}
\affiliation{Research Center for Electron Photon Science, Tohoku University, Sendai 980-8578}
\affiliation{University of Science and Technology of China, Hefei 230026}
\affiliation{Seoul National University, Seoul 151-742}
\affiliation{Soongsil University, Seoul 156-743}
\affiliation{Sungkyunkwan University, Suwon 440-746}
\affiliation{School of Physics, University of Sydney, NSW 2006}
\affiliation{Tata Institute of Fundamental Research, Mumbai 400005}
\affiliation{Excellence Cluster Universe, Technische Universit\"at M\"unchen, 85748 Garching}
\affiliation{Toho University, Funabashi 274-8510}
\affiliation{Tohoku Gakuin University, Tagajo 985-8537}
\affiliation{Tohoku University, Sendai 980-8578}
\affiliation{Department of Physics, University of Tokyo, Tokyo 113-0033}
\affiliation{Tokyo Institute of Technology, Tokyo 152-8550}
\affiliation{Tokyo Metropolitan University, Tokyo 192-0397}
\affiliation{Tokyo University of Agriculture and Technology, Tokyo 184-8588}
\affiliation{University of Torino, 10124 Torino}
\affiliation{CNP, Virginia Polytechnic Institute and State University, Blacksburg, Virginia 24061}
\affiliation{Wayne State University, Detroit, Michigan 48202}
\affiliation{Yamagata University, Yamagata 990-8560}
\affiliation{Yonsei University, Seoul 120-749}
  \author{C.~P.~Shen}\affiliation{Beihang University, Beijing 100191} 
  \author{C.~Z.~Yuan}\affiliation{Institute of High Energy Physics, Chinese Academy of Sciences, Beijing 100049} 
  \author{A.~Sibidanov}\affiliation{School of Physics, University of Sydney, NSW 2006} 
  \author{P.~Wang}\affiliation{Institute of High Energy Physics, Chinese Academy of Sciences, Beijing 100049} 
  \author{K.~Hayasaka}\affiliation{Kobayashi-Maskawa Institute, Nagoya University, Nagoya 464-8602} 
  \author{X.~L.~Wang}\affiliation{CNP, Virginia Polytechnic Institute and State University, Blacksburg, Virginia 24061} 
  \author{I.~Adachi}\affiliation{High Energy Accelerator Research Organization (KEK), Tsukuba 305-0801} 
  \author{H.~Aihara}\affiliation{Department of Physics, University of Tokyo, Tokyo 113-0033} 
  \author{D.~M.~Asner}\affiliation{Pacific Northwest National Laboratory, Richland, Washington 99352} 
  \author{T.~Aushev}\affiliation{Institute for Theoretical and Experimental Physics, Moscow 117218} 
% ...
\collaboration{The Belle Collaboration}

\begin{abstract}

Using data samples of 89~fb$^{-1}$, 703~fb$^{-1}$, and 121~fb$^{-1}$
collected with the Belle detector at the KEKB asymmetric-energy $e^+e^-$ collider
at center-of-mass energies 10.52 GeV, 10.58 GeV, and 10.876 GeV, respectively,
we study the exclusive reactions
$e^+e^- \to \omega\pi^0$, $K^{\ast}(892)\bar{K}$, and
$K_2^{\ast}(1430)\bar{K}$ (Charge-conjugate modes are included implicitly).
Significant signals of $\omega\pi^0$,
$K^{\ast}(892)^0\bar{K}^0$, and $K_2^{\ast}(1430)^-K^+$ are
observed for the first time at these energies, and
the energy dependencies of the cross sections are presented. On the other hand, no significant excesses for
$K^{\ast}(892)^-K^+$ and $K_2^{\ast}(1430)^0 \bar{K}^0$ are found,
and we set limits on the cross section ratios $R_{\rm VP} =
\frac{\sigma_B(e^+e^-\to K^{\ast}(892)^0\bar K^0)} {\sigma_B(e^+e^-\to
K^{\ast}(892)^-K^+ )}>$ 4.3, 20.0, and 5.4, and $R_{\rm TP}
= \frac{\sigma_B(e^+e^-\to K_2^{\ast}(1430)^0\bar K^0)} {\sigma_B(e^+e^-\to
K_2^{\ast}(1430)^-K^+)}<$ 1.1, 0.4, and 0.6, for center-of-mass  energies
of 10.52 GeV, 10.58 GeV, and 10.876~GeV, respectively,  at the 90\% C.L.

\end{abstract}


\maketitle


Large data samples collected at the B-factories provide an
opportunity to explore rare two-meson production in $\EE$
annihilation, which allows us to investigate the energy dependence
of various meson form factors and shed light on hadron structure
and hence the strong
interaction. These studies also supply information on the wave function
of hadrons.

For a center-of-mass (CM) energy $\sqrt{s}$ much larger than resonance masses, one expects that
the proportions of the cross sections of $\omega\piz:K^{\ast
}(892)^0{\bar K}^0:K^{\ast}(892)^- K^+$  production equal 9:8:2~\cite{charged}
if SU(3) flavor symmetry is exact. However, this relation was found to be
violated severely at $\sqrt{s}=3.67$ GeV and 3.773~GeV by the CLEO
experiment~\cite{cleo-kstrk}, with the $\omega\piz$ cross sections
smaller than those of the $K^{\ast}(892)^0{\bar K}^0$, and the
ratio $R_{\rm VP}=\frac{\sigma_B(e^+e^-\to K^{\ast}(892)^0\bar
K^0)}{\sigma_B(e^+e^-\to K^{\ast}(892)^-K^+)}$ greater than 9
and 33 at $\sqrt{s}=3.67$ GeV and 3.773~GeV, respectively, at the 90\%
confidence level (C.L.)~\cite{ourowncalculation}.

By taking into account  SU(3)$_{\rm f}$ symmetry breaking
and the transverse momentum distribution of  partons in the
light cone wave functions of  mesons, a pQCD
calculation~\cite{LC} can reproduce most of the CLEO
measurements with reasonable input parameters, and the corresponding
cross sections at $\sqrt{s}=10.58$~GeV are predicted.
The calculation predicts $R_{\rm VP}=6.0$, which is
far below the CLEO lower limits and may indicate deficiencies
in the model assumptions. The same calculation
also predicts that the cross sections of $\EE\to$ vector-pseudoscalar (VP) vary as
$1/s^3$ rather than $1/s^2$ in Ref.~\cite{gerard} or $1/s^4$ in
Refs.~\cite{stanrule,Chernyak,Likhoded}; this can also be
tested by combining the measurements from CLEO and the
B-factories.
At Belle, the cross sections of
$e^+e^- \to \phi\eta,~\phi\eta',~\rho\eta,~\rho\eta'$
have been measured at $\sqrt{s}=10.58$ GeV; however, no definite
conclusion about the energy dependence of
$e^+e^- \rightarrow VP$ can be drawn~\cite{belle-VP}.

In the quark model, the tensor states $K_2^{\ast}(1430)$ have the
same quark content as the vector states $K^{\ast}(892)$; thus, one may
naively expect the same ratio between the neutral and charged
$K_2^{\ast}(1430)\bar{K}$ production in $\EE$ annihilation as in the VP
case, i.e.,  $R_{\rm TP}=\frac{\sigma_B(e^+e^-\to K_2^{\ast}(1430)^0\bar
K^0)}{\sigma_B(e^+e^-\to K_2^{\ast}(1430)^-K^+)}=R_{\rm VP}$.
This has never been tested.



In this paper, we report the cross sections of the exclusive
reactions $\EE\to \omega \pi^0$, $K^{\ast}(892) \bar{K}$, and
$K_2^{\ast}(1430) \bar{K}$, based on data samples of 89~fb$^{-1}$, 703~fb$^{-1}$,
and 121~fb$^{-1}$ collected at $\sqrt{s}=$10.52, 10.58 ($\Upsilon(4S)$ peak), and
10.876~GeV ($\Upsilon(5S)$ peak), respectively.
The data were collected with the Belle
detector~\cite{Belle} operating at the KEKB asymmetric-energy
$\EE$ collider~\cite{KEKB}. The final states are $\pp\piz\piz$ and
$K^0_SK^+\pi^-$, in which the $K^0_S$ is reconstructed from $\pp$.
The generator {\sc mcgpj}, developed according to the
calculations in Ref.~\cite{mcjpg}, is used to generate Monte Carlo (MC)
events with the exact next-to-leading order radiative corrections
applied to all the studied processes. Generic  $\EE \to
u\bar{u}/d\bar{d}/s\bar{s}$ MC events, produced using {\sc
pythia}~\cite{pythia}, are used to check background
contributions.

The Belle detector is described in detail elsewhere~\cite{Belle}.
It is a large-solid-angle magnetic spectrometer that consists of a
silicon vertex detector (SVD), a 50-layer central drift chamber
(CDC), an array of aerogel threshold Cherenkov counters (ACC), a
barrel-like arrangement of time-of-flight scintillation counters
(TOF), and an electromagnetic calorimeter comprised of CsI(Tl)
crystals (ECL) located inside a superconducting solenoid coil that
provides a 1.5~T magnetic field. An iron flux return located
outside of the coil is instrumented to detect $K_L^0$ mesons and
to identify muons (KLM).


For each charged track except those from $\ks$ decays, the impact
parameters perpendicular to and along the beam direction with
respect to the interaction point are required to be less than
0.5~cm and 4~cm, respectively, and the transverse momentum must
exceed 0.1~GeV/$c$ in the laboratory frame. Well-measured charged
tracks are selected and the numbers of such charged tracks are two
for the $\pp \pi^0 \pi^0$ final state and four for the $\ks K^+
\pi^-$ final state. For each charged track, we combine information
from several detector subsystems to form a likelihood
$\mathcal{L}_i$ for each particle species~\cite{pid}. A track with
$\mathcal{R}_K = \frac{\mathcal{L}_K} {\mathcal{L}_K +
\mathcal{L}_\pi}> 0.6$ is identified as a kaon, while a track with
$\mathcal{R}_K<0.4$ is treated as a pion. With this selection, the
kaon (pion) identification efficiency is about 85\% (89\%), while
6\% (9\%) of kaons (pions) are misidentified as pions (kaons). For
electron identification, the likelihood ratio is defined as
$\mathcal{R}_e = \frac{\mathcal{L}_e}
{\mathcal{L}_e+\mathcal{L}_x}$, where $\mathcal{L}_e$ and
$\mathcal{L}_x$ are the likelihoods for electron and non-electron,
respectively. These are determined using the ratio of the energy
deposited in the ECL to the momentum measured in the SVD and CDC,
the shower shape in the ECL, position matching between
the charged track trajectory and the cluster position in the ECL,
hit information from the ACC, and specific ionization ($dE/dx$) information in the
CDC~\cite{EID}. For muon identification, the likelihood ratio is
defined as $\mathcal{R}_\mu = \frac{\mathcal{L}_\mu}
{\mathcal{L}_\mu+\mathcal{L}_\pi+\mathcal{L}_K}$, where
$\mathcal{L}_\mu$, $\mathcal{L}_\pi$, and $\mathcal{L}_K$ are the
likelihoods for muon, pion, and kaon, respectively. These are
based on track matching quality and penetration depth of
associated hits in the KLM~\cite{MUID}.

Except for the $\pp$ pair from $\ks$ decay, all charged tracks
are required to be positively identified as pions or kaons.
The requirements $\mathcal{R}_\mu< 0.95$ and $\mathcal{R}_e<0.95$
for the charged tracks remove 9.3\% of the backgrounds for
$K_S^0 K^+ \pi^-$ with negligible loss in efficiency.

For $\ks$ candidates decaying into $\pp$ in the $\ks K^+ \pi^-$
mode, we require that the invariant mass of the $\pp$ pair lie
within a $\pm 8$~MeV/$c^2$ interval around the $\ks$ nominal mass,
which contains around 95\% of the signal according to MC
simulation, and that the pair have a displaced vertex and flight
direction consistent with a $\ks$ originating from the
IP~\cite{ks}.

An energy cluster in the electromagnetic calorimeter is
reconstructed as a photon if it does not match the extrapolated
position of any charged track.
A $\pi^0$ candidate
is reconstructed from a pair of photons whose energies exceed 100 MeV in the laboratory frame. We perform a
mass-constrained fit to the selected $\pi^0$ candidate and require
$\chi^2<15$.
To suppress background from the Initial-State-Radiative (ISR) process $\EE \to
\gamma_{\rm ISR}\omega \to \gamma_{\rm ISR} \pp \pi^0$,
the requirement of $|(E_1-E_2)/(E_1+E_2)|<0.65$ is imposed
for the primary $\pi^0$ of $\EE \to \omega \pi^0$, where $E_1$ and $E_2$ are the energies
 in the laboratory frame
of the photons forming the higher-momentum $\pi^0$ candidate.

\begin{figure}[htbp]
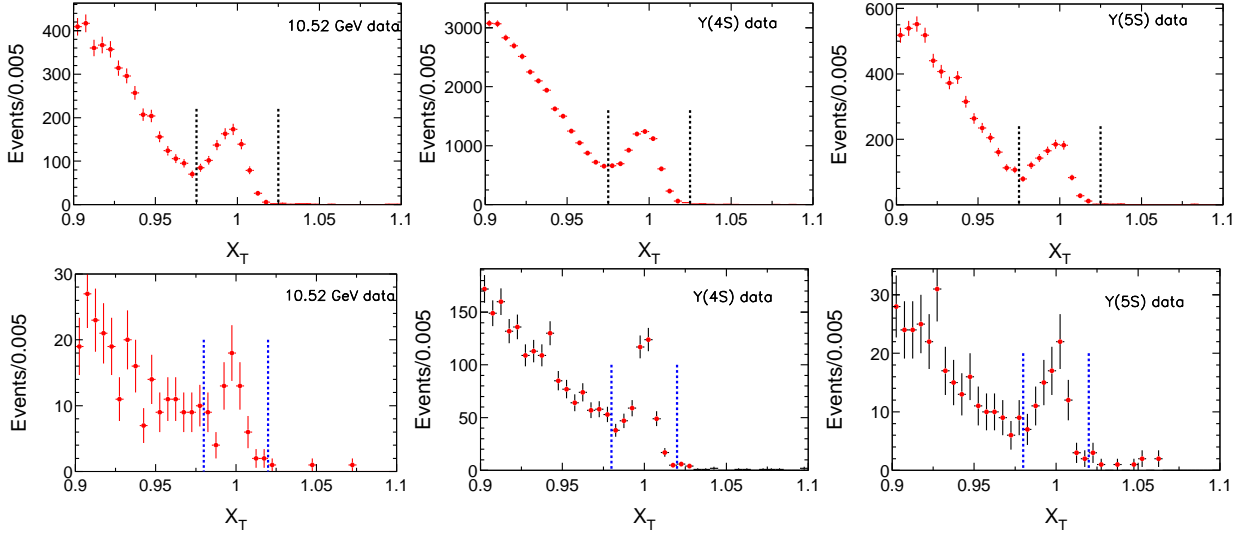

\includegraphics[height=3.5cm]{fig1a.epsi}
\includegraphics[height=3.5cm]{fig1b.epsi}
\includegraphics[height=3.5cm]{fig1c.epsi}
\includegraphics[height=3.5cm]{fig1d.epsi}
\includegraphics[height=3.5cm]{fig1e.epsi}
\includegraphics[height=3.5cm]{fig1f.epsi}
\caption{\label{ecms} The scaled total energy $X_T$ distributions
for the selected $\EE \to \pp \pi^0 \pi^0$ (top
row) and $\ks K^+ \pi^-$ (bottom row) candidate events
from the $\sqrt{s}=10.52$ GeV,
10.58 GeV, and 10.876~GeV data samples.
The signal region is between the dotted lines.}
\end{figure}

We define an energy conservation variable $X_{T} = \Sigma_h
E_h/\sqrt{s}$, where $E_h$ is the energy of the final-state
particle $h$ in the $\EE$ CM frame. For the signal candidates,
$X_T$ should be around 1. After the application of all the above
selection requirements, Fig.~\ref{ecms} shows the $X_T$
distributions for the final candidate events of $\EE \to \pp \pi^0
\pi^0$ (top row) and $\ks K^+ \pi^-$ (bottom row) from the
$\sqrt{s}=10.52$ GeV, 10.58 GeV, and 10.876~GeV data samples,
respectively. Clear $\EE \to \pp \pi^0 \pi^0$ and $\ks K^+ \pi^-$
signals are observed. We require $|X_T-1|<0.025$ for $\pp \pi^0
\pi^0$ and $|X_T-1|<0.02$ for $\ks K^+ \pi^-$, as indicated by
the dotted lines in Fig.~\ref{ecms}.

\begin{figure}[htbp]
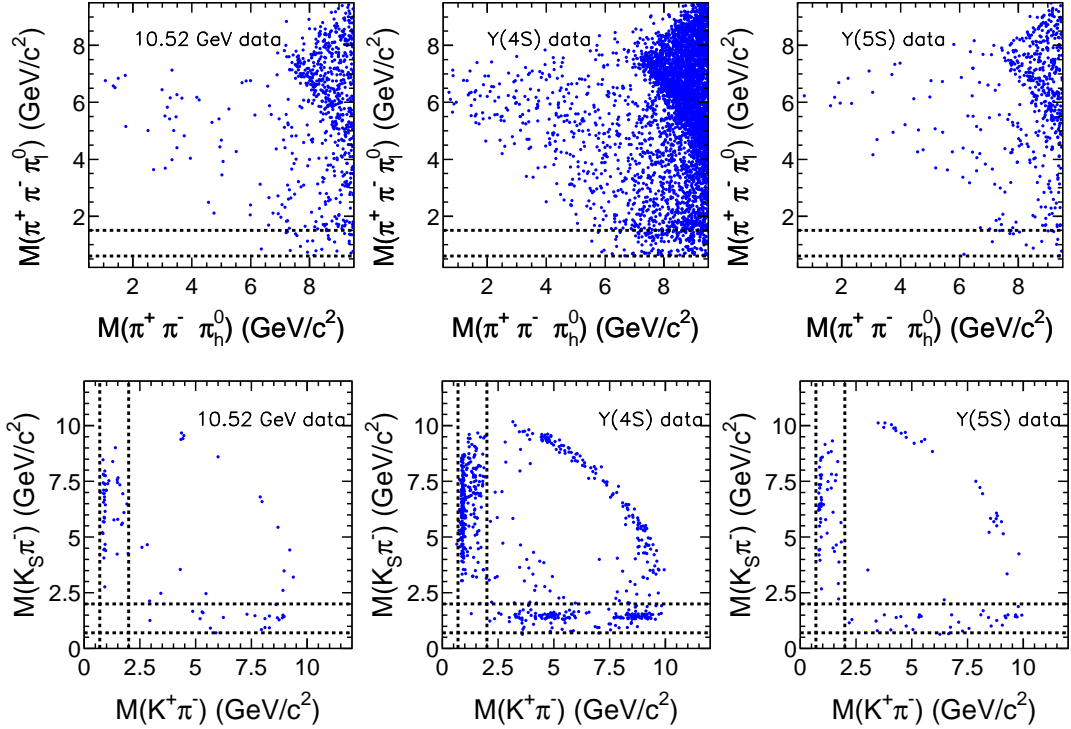

\includegraphics[height=4.6cm]{fig2a.epsi}
\includegraphics[height=4.6cm]{fig2b.epsi}
\includegraphics[height=4.6cm]{fig2c.epsi}\vspace{0.4cm}
\includegraphics[height=4.6cm]{fig2d.epsi}
\includegraphics[height=4.6cm]{fig2e.epsi}
\includegraphics[height=4.6cm]{fig2f.epsi}
\caption{\label{dalitz} Distributions of $M(\pp \pi^0_l)$ versus $M(\pp\pi^0_h)$
for the $\pp \pi^0 \pi^0$ (top row) and $M(\ks \pi^-)$ versus $M(K^+ \pi^-)$
for the $\ks K^+ \pi^-$ (bottom row) final states
from the $\sqrt{s}=10.52$ GeV, 10.58 GeV, and 10.876~GeV data samples.
In the $\pp \pi^0 \pi^0$ panels, $\pi^0_h$ and $\pi^0_l$
represent the pions with  higher and lower momentum in the
laboratory system, respectively. The events between the dotted lines
will be selected to search for $\omega$, $K^{\ast}$ and $K_2^{\ast}$ signals.}
\end{figure}

The distributions of $M(\pp \pi^0_l)$ versus $M(\pp\pi^0_h)$ for
the $\pp \pi^0_h \pi^0_l$ final state and $M(\ks \pi^-)$ versus $M(K^+ \pi^-)$
for the $\ks K^+ \pi^-$ final state are shown in Fig.~\ref{dalitz}.
Here, $\pi^0_h$ and $\pi^0_l$ represent the $\pi^0$ candidates with
higher and lower momentum, respectively, in the laboratory system.
According to MC-simulated $\EE\to \omega \pi^0$
signal events, most of the $\pi^0$s ($>97\%$) from $\omega$ decays
have lower momentum and there is only one $\pp\pi^0$ combination
in the $\omega$ mass region. In the $\ks K^+ \pi^-$ mode,
we see clearly the intermediate states $K^{\ast}(892)\bar{K}$, $K_2^{\ast}(1430)\bar{K}$ and
possibly $a_2(1320)\pi$.

\begin{figure}[htbp]
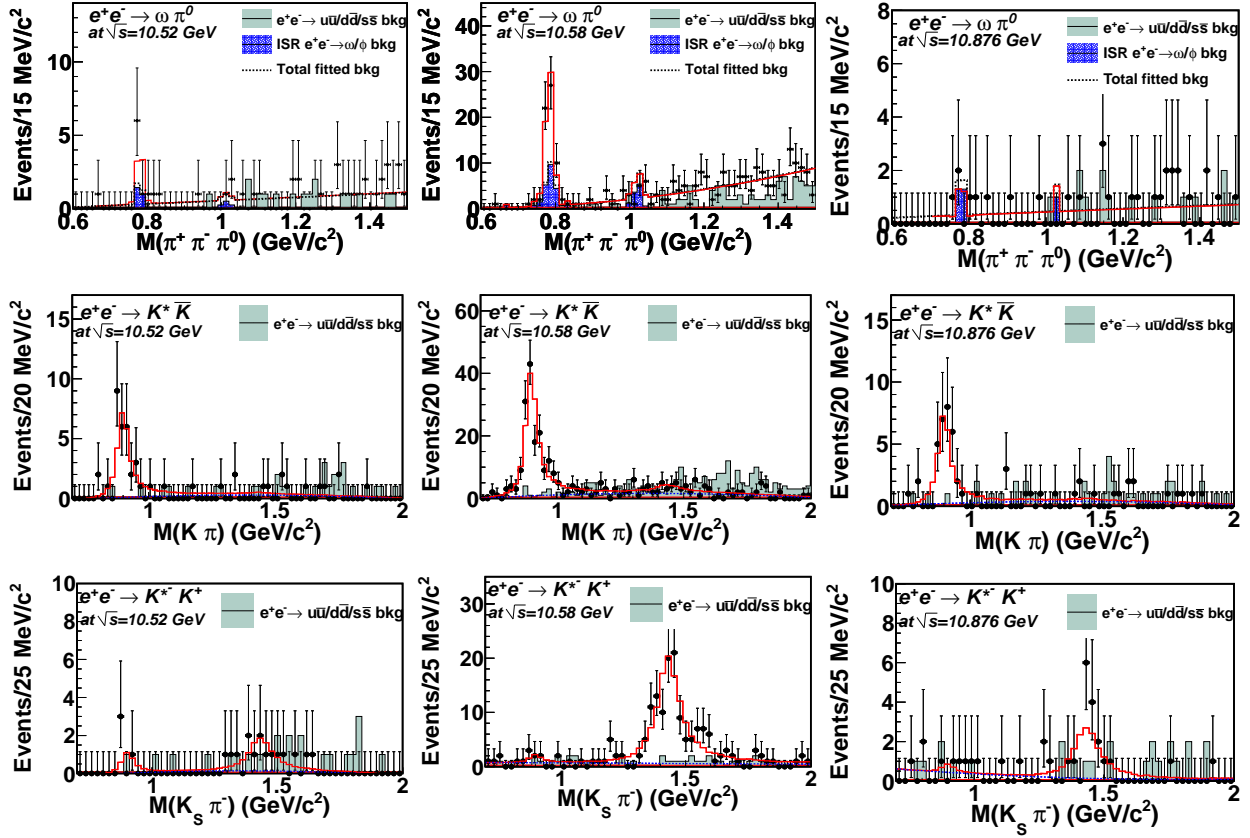

\includegraphics[height=5.3cm,angle=-90]{fig3a.epsi}
\includegraphics[height=5.3cm,angle=-90]{fig3b.epsi}
\includegraphics[height=5.5cm,angle=-90]{fig3c.epsi}\vspace{0.25cm}
\includegraphics[height=5.3cm,angle=-90]{fig3d.epsi}
\includegraphics[height=5.3cm,angle=-90]{fig3e.epsi}
\includegraphics[height=5.5cm,angle=-90]{fig3f.epsi}\vspace{0.25cm}
\includegraphics[height=5.3cm,angle=-90]{fig3g.epsi}
\includegraphics[height=5.3cm,angle=-90]{fig3h.epsi}
\includegraphics[height=5.5cm,angle=-90]{fig3i.epsi}
\caption{\label{vt-fit} The fits to the $\pp \pi^0$ (top row),
$K^+ \pi^-$ (middle row) and $\ks \pi^-$ (bottom row) invariant
mass distributions for the $\omega$, $K^{\ast}(892)$, and $K_2^{\ast}(1430)$
meson candidates from $\EE \to \pp \pi^0 \pi^0$ and $\ks K^+
\pi^-$ events from the $\sqrt{s}=10.52$ GeV, 10.58 GeV, and 10.876~GeV data
samples. The solid lines show the results of the
fits described in the text, the dotted curves show the total
background estimates, the dark shaded histograms are from the
normalized ISR backgrounds $\EE \to \gamma_{\rm ISR} \omega/\phi \to \gamma_{\rm ISR} \pp \pi^0$ and
the light shaded histograms are from the normalized $\EE \to
u\bar{u}/d\bar{d}/s\bar{s}$ backgrounds. The dotted curves are not significantly seen in the signal
 regions due to low background level.}
\end{figure}

For the selected events, Fig.~\ref{vt-fit} shows the $\pp \pi^0$,
$K^+ \pi^-$, and $\ks \pi^-$ invariant mass distributions for the
$\pp\piz\piz$ and $\ks K^+ \pi^-$ final states from the
$\sqrt{s}=10.52$ GeV, 10.58 GeV, and 10.876~GeV data samples.
For charge-conjugate modes
the numbers of selected candidate events are consistent within one standard deviation.
The dots with error bars are from data and the
light shaded histograms are from the normalized $\EE \to
u\bar{u}/d\bar{d}/s\bar{s}$ backgrounds. In the $\pp \pi^0$
invariant mass distributions, the dark shaded histograms in the
$\omega$ and $\phi$ mass regions are from the normalized $\EE
\to \gamma_{\rm ISR} \omega/\phi \to \gamma_{\rm ISR} \pp \pi^0$ backgrounds. In the normalization,
the expected ISR events are calculated with $N^{\rm prod}=\lum \times
\sigma^{\rm prod}$, where $\lum$ is the integrated luminosity and
$\sigma^{\rm prod}$ is the production cross section. The
production cross sections are calculated to be $\sigma^{\rm
prod}(\EE \to \gamma_{\rm ISR} \omega)=15.1$~pb, 14.9~pb, and
14.2~pb, and $\sigma^{\rm prod}(\EE \to \gamma_{\rm ISR} \phi)
=25.4$~pb, 25.2~pb, and 23.9~pb, for $\sqrt{s}=10.52$ GeV, 10.58 GeV, and
10.876~GeV, respectively~\cite{QWG}. ISR MC events of $\EE \to
\gamma_{\rm ISR} \omega/\phi \to \gamma_{\rm ISR} \pp \pi^0$ are simulated using the {\sc phokhara}
generator~\cite{phokahara}, which simulates ISR process at the
next-to-leading order accuracy. In the $K^+ \pi^-$ and $\ks \pi^-$
invariant mass distributions, we observe clear $K^{\ast}(892)^0$ and
$K_2^{\ast}(1430)^-$ signals, while almost no signals for $K_2^{\ast
}(1430)^0$ and $K^{\ast}(892)^-$ can be seen.

We perform unbinned maximum likelihood fits to these mass distributions,
as shown in Fig.~\ref{vt-fit}. The signal shapes of $\omega$,
$K^{\ast}(892)$, and $K_2^{\ast}(1430)$ are
obtained directly from MC simulated signal samples~\cite{mass-reso}.
The combinatorial backgrounds are modeled  by a
second-order Chebyshev polynomial and the additional normalized backgrounds from  $\EE \to
\gamma_{\rm ISR} \omega/\phi \to \gamma_{\rm ISR} \pp \pi^0$ are fixed in the
$\pp \pi^0$ mass spectrum fit.
The fitted results are shown in Fig.~\ref{vt-fit} and listed in
Table~\ref{summary}.

The significances and the upper limits listed in Table~\ref{summary}
are obtained by evaluating the likelihood profile. To take into account the systematic
uncertainty, we convolve the likelihood function with a Gaussian whose width
equals the total systematic uncertainty. The significance is obtained by
comparing the likelihood values at maximum and at zero signal yield using
$\sqrt{-2\ln(\mathcal{L}_0/\mathcal{L}_{\rm max})}$.
The upper limit $N^{\rm UL}_{\rm sig}$  on $N_{\rm sig}$ at $90\%$ C.L. is obtained
by integrating the likelihood function from zero to the bound that
gives $90\%$ of the total area.

The observed cross section is determined according to the formula
$\sigma^{\rm obs} = \frac{N}{L\, B_{V/T}\, B_P\, \epsilon}, $ where $N$ is
the signal yield, $L$ is the integrated luminosity, $B_{\rm V/T}$
and $B_{\rm P}$ are the branching fractions of the corresponding
decay channels of the vector/tensor and pseudoscalar mesons
including secondary branching fractions to reconstructed final states,
respectively, and $\epsilon$ is the corresponding detection
efficiency. The Born cross section is written as
$\sigma_B=\frac{\sigma^{\rm obs}|1-\Pi(s)|^2}{(1+\delta)}$,
where $1+\delta$ is the radiative correction factor and $|1-\Pi(s)|^2$ is the vacuum polarization
factor.
The radiative correction factors $1+\delta$ are 0.89, 0.88, and 0.88 for $\omega \pi^0$, $K^{\ast}(892)\bar{K}$,
and $K_2^{\ast}(1430)\bar{K}$, respectively, calculated with a
limit on the energy of the radiated photon
of 0.5~GeV~\cite{mcjpg}; the values of $|1-\Pi(s)|^2$ are 0.931, 0.930, and 0.929~\cite{vp}
for $\sqrt{s}=10.52$ GeV, 10.58 GeV and 10.876 GeV, respectively.

\begin{table}[htbp]
\caption{Results for the Born  cross sections, where $N_{\rm
sig}$ is the number of fitted signal events, $N^{\rm UL}_{\rm
sig}$ is the upper limit on the number of signal events,
$\epsilon$ is the efficiency, $\Sigma$ is the signal
significance, $\sigma_B$ is the Born  cross section,
$\sigma_B^{\rm UL}$ is the upper limit on the Born  cross
section. All the upper limits are given at the 90\% C.L. The
first uncertainty in $\sigma_B$ is statistical,
and the second systematic.} \label{summary}
\begin{center}
\begin{tabular}{c|ccccccc}
\hline\hline
Channel & $\sqrt{s}$ (GeV) &$N_{\rm sig}$ & $N^{\rm UL}_{\rm sig}$& $\epsilon$ (\%) & $\Sigma$ ($\sigma$) & $\sigma_B$ (fb)
        &$\sigma_B^{\rm UL}$ (fb)   \\ \hline
$\omega \pi^0$& 10.52 &$4.1^{+3.3}_{-2.6}$& 9.9& 1.25& 1.6&$4.53^{+3.64}_{-2.88}\pm 0.50$ & 11 \\
              & 10.58 &$38.8^{+8.3}_{-7.6}$& ---& 1.10& 6.7&$6.01^{+1.29}_{-1.18}\pm0.57$ &--- \\
              & 10.876 &$-0.7^{+2.9}_{-2.1}$& 7.0& 1.07& ---&$-0.68^{+2.71}_{-1.97}\pm 0.20$ & 6.5 \\
\hline \rule{0mm}{0.4cm}
$K^{\ast}(892)^0\bar{K}^0$& 10.52 &$34.6^{+6.9}_{-6.1}$&---&16.49& 7.4&$10.77^{+2.15}_{-1.90}\pm0.77$&---\\
                    & 10.58 &$187\pm 17$&---&16.30& $>$10&$7.48\pm0.67\pm 0.51$&---\\
                    & 10.876&$34.6^{+7.5}_{-6.7}$&---&17.25& 7.2&$7.58^{+1.64}_{-1.47}\pm0.63$&---\\
\hline \rule{0mm}{0.4cm}
$K^{\ast}(892)^-K^+$& 10.52 & $4.6^{+3.6}_{-2.7}$ & 9.3 & 20.40 & 1.4 & $1.14^{+0.90}_{-0.67}\pm0.15$ & 2.3 \\
              & 10.58 & $5.9^{+4.7}_{-3.8}$ & 14 & 21.03 & 1.5 & $0.18^{+0.14}_{-0.12}\pm 0.02$ & 0.4\\
              & 10.876 & $1.6^{+3.9}_{-3.0}$ & 8.5 & 21.29 & 0.3 & $0.28^{+0.68}_{-0.52}\pm0.10$ & 1.5  \\
\hline \rule{0mm}{0.4cm}
$K_2^{\ast}(1430)^0\bar{K}^0$&10.52 &$1.3^{+4.3}_{-3.9}$ & 6.8 & 17.63 & 0.3 & $0.76^{+2.53}_{-2.26}\pm 0.14$ & 4.0 \\
                      &10.58 & $21^{+11}_{-10}$ & 40 & 16.71 & 2.1 & $1.65^{+0.86}_{-0.78}\pm 0.27$ & 3.1 \\
                      &10.876& $1.0^{+4.5}_{-3.7}$ & 8.9 & 19.02 & 0.2 & $0.38^{+1.79}_{-1.47}\pm0.07$ & 3.5   \\
\hline \rule{0mm}{0.4cm}
$K_2^{\ast}(1430)^-K^+$&10.52 & $12.0^{+6.2}_{-5.8}$ & 21 & 20.36 & 2.1 & $6.06^{+3.13}_{-2.93}\pm1.34$ & 11  \\
                      &10.58 & $129\pm 15$ & --- & 20.17 & $>$10 & $8.36\pm0.95\pm0.62$ & --- \\
                      &10.876& $17.6^{+5.3}_{-4.6}$ &--- & 21.50 & 4.5 & $6.20^{+1.86}_{-1.63}\pm0.64$ & ---   \\
\hline\hline
\end{tabular}
\end{center}
\end{table}


There are several sources of systematic uncertainties for the cross
section measurements. The uncertainty in the tracking efficiency
for tracks with angles and momenta characteristic of signal events
is about 0.35\% per track and is additive. The uncertainty due to
particle identification efficiency is 1.7\% with an efficiency
correction factor of 0.98 for each pion and is 1.6\% with an
efficiency correction factor 0.97 for each kaon. The uncertainty
in selecting $\pi^0$ is estimated using a control sample of $\tau^- \to
\pi^- \pi^0 \nu_{\tau}$. We introduce a 2.2\% systematic
uncertainty with efficiency correction factors of 0.94 for a low momentum
$\pi^0$ and 0.97 for a high momentum one.
In the $\ks K^+ \pi^-$
mode, the $K_S^0$ reconstruction systematic uncertainty is estimated by comparing the
ratio of the $D^+\to K_S^0\pi^+$ and $D^+\to K^-\pi^+\pi^+$ yields
with the MC expectations;
the difference between data and MC
simulation is less than 4.9\%~\cite{ks-error}.
Uncertainties on the branching fractions of the intermediate states are taken from the
PDG listings~\cite{PDG}. According to MC simulation, the trigger
efficiency is greater than 99\% so the corresponding uncertainty
is neglected. We estimate the systematic uncertainties associated
with the fitting procedure by changing the shape of the background
and the range of the fit and taking the differences in
the fitted results, which are 1.0\%-32\% depending on the final
state particles, as systematic uncertainties. The uncertainty due to
limited MC statistics is at most 2.4\%.
The form factor dependence on $s$ is assumed to be $\frac{1}{s}$
in the {\sc mcgpj} generator for the nominal results. The differences in the efficiency
compared to the assumption of $\frac{1}{s^2}$ dependence for the form
factor are taken as the systematic uncertainties due to the generator uncertainty,
which are 1.5\%, 0.9\%, and 0.9\% for the $\omega \pi^0$,
$K^{\ast}(892) \bar{K}$, and $K_2^{\ast}(1430) \bar{K}$, respectively.
We take 2\% systematic uncertainty due to the uncertainty of
the effect of soft and virtual photon emission in the generator~\cite{mcjpg}.
The efficiency differences are 0.7\% and 1.3\% for $\omega \pi^0$
and $\ks K^+ \pi^-$ final states, respectively, when including or excluding final state radiation~\cite{photos};
these are included into the uncertainty of the generator.
Finally, the total
luminosity is determined using wide angle Bhabha events with
1.4\% precision.
Assuming that all of these systematic uncertainty sources are
independent, the total systematic uncertainty is 6.8\%-33\%, depending on
the final state, as shown in Table~\ref{totalsys}.

\begin{table}[htbp]
\caption{Relative systematic uncertainties (\%) on the cross section.
For the fit uncertainty and the total systematic uncertainty, the three values
separated by slashes are for the
CM energies 10.52 GeV, 10.58 GeV, and 10.876~GeV, respectively.} \label{totalsys}{\footnotesize
\begin{tabular}{l | c | c c | c c }
\hline Source & $\omega \pi^0$ & $K^{\ast}(892)^0 \bar{K}^0$  & $K^{\ast}(892)^- K^+$& $K_2^{\ast}(1430)^0 \bar{K}^0$  & $K_2^{\ast}(1430)^- K^+$
 \\\hline
 Tracking & 0.7 & 0.7 & 0.7 & 0.7  & 0.7  \\
 PID      & 3.4 & 3.3 & 3.3 & 3.3 & 3.3  \\
 $\pi^0$ selection & 4.4 & --- & ---  & --- & ---  \\
 $\ks$ selection   & --- & 4.9 & 4.9 &  4.9 & 4.9  \\
  Branching fractions & 0.8 & 0.1 & 0.1 & 2.4 & 2.4  \\
 Fit uncertainty & 8.8/6.6/28 & 2.4/1.0/4.9  & 11/8.2/32  & 16/14/15 & 21/2.2/7.4  \\
 MC statistics & 2.4  & 0.8 & 0.8 & 0.7 & 0.7  \\
 Generator& 2.6 & 2.6  & 2.6 & 2.6 & 2.6  \\
 Luminosity & 1.4 & 1.4 & 1.4 & 1.4 & 1.4  \\
 \hline
 Sum in quadrature &11/9.5/29 & 7.1/6.8/8.3  & 13/11/33  & 18/16/17 & 22/7.4/11  \\\hline
\end{tabular}}
\end{table}


Table~\ref{summary} shows the results for the measured Born
cross sections including the upper limits at 90\% C.L. for the
channels with a signal significance of less than 3$\sigma$.
These are the first measurements of the cross sections and upper limits
at CM energies 10.52 GeV, 10.58 GeV, and 10.876~GeV.
The measured cross sections of
$\EE \to \omega \pi^0$ and $K^{\ast}(892)^0 \bar{K}^0$ at
$\sqrt{s}=10.58$~GeV are consistent within errors with the
theoretical predictions that range from $(4.1^{+0.5}_{-0.3})$ fb to
$(5.2^{+0.4}_{-0.3})$ fb for $\omega \pi^0$ and from
$(5.6^{+0.2}_{-0.4})$ fb to $(7.1\pm0.4)$ fb for
$K^{\ast}(892)^0 \bar{K}^0$ in Ref.~\cite{LC}.
In contrast, we do not observe a significant signal for $\EE \to K^{\ast}(892)^-
K^+$ and the upper limit of the cross section at 10.58~GeV
is much lower than the prediction from the same
calculation~\cite{LC}. The
measured cross section of $\EE \to \omega \pi^0$ is much smaller
than the calculated value of about 240 fb using the theoretical formulae in
Ref.~\cite{gerard}.

\begin{figure}[htbp]
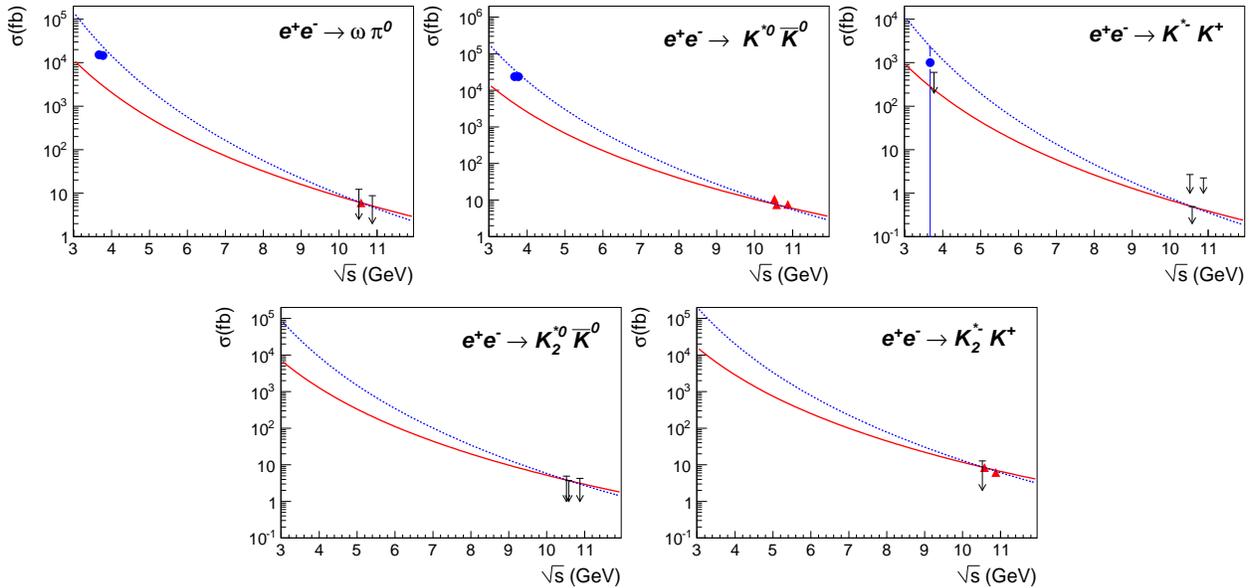

\includegraphics[height=5.4cm,angle=-90]{fig4a.epsi}
\includegraphics[height=5.4cm,angle=-90]{fig4b.epsi}\vspace{0.25 cm}
\includegraphics[height=5.4cm,angle=-90]{fig4c.epsi}
\includegraphics[height=5.4cm,angle=-90]{fig4d.epsi}
\includegraphics[height=5.4cm,angle=-90]{fig4e.epsi}
\caption{\label{cs-k2str} The cross sections
for  $\EE \to \omega
\pi^0$, $K^{\ast}(892) \bar{K}$, and $K_2^{\ast}(1430) \bar{K}$.
The data at $\sqrt{s}=10.52$ GeV, 10.58 GeV, and 10.876~GeV
are from our measurements. The data at $\sqrt{s}=3.67$ GeV and 3.77~GeV,
where shown, are from  CLEO
measurement~\cite{cleo-kstrk}.
Here, the uncertainties are the sum of the statistical
and systematic uncertainties in quadrature.
Upper limits are shown by the arrows. The solid
line corresponds to a $1/s^3$ dependence and the dashed line
to a $1/s^4$ dependence; the curves pass through the measured cross
section at $\sqrt{s}=10.58$~GeV.}
\end{figure}

Figure~\ref{cs-k2str} shows the cross sections measured in our experiment
at $\sqrt{s}=10.52$ GeV, 10.58 GeV, and 10.876~GeV for
$\EE \to \omega \pi^0$, $K^{\ast}(892) \bar{K}$ and $K_2^{\ast}(1430)
\bar{K}$, where the uncertainties are the sum in quadrature  of the statistical
and systematic uncertainties.
Since the signal significance is greater than $5\sigma$
for $\EE \to K^{\ast}(892)^0\bar{K}^0$ at all energies and
for $\EE \to \omega \pi^0$ at $\sqrt{s}=10.58$ GeV, we fit the
$1/s^{n}$ dependence of the cross sections
to our data and those from CLEO at $\sqrt{s}=3.67$ GeV and 3.77~GeV~\cite{cleo-kstrk}.
The fit gives $n=3.83\pm0.07$ and
$3.75\pm0.12$ for $\EE \to K^{\ast}(892)^0\bar{K}^0$
and $\omega \pi^0$~\cite{ynso}, respectively.
These differ significantly from the $1/s^2$~\cite{gerard}
or $1/s^3$~\cite{LC} predictions and agree with
$1/s^4$~\cite{stanrule,Chernyak,Likhoded} within $2.5\sigma$.
For other channels,
no definite conclusion can be drawn from
current results due to the large uncertainties.

In all the above discussions, we neglect possible small contributions from $\Upsilon(4S)$ and
$\Upsilon(5S)$ resonance decays in the measured Born cross sections at $\sqrt{s}=10.58$ GeV
and 10.876 GeV. Since the signal significance exceeds $5\sigma$
for the $K^{\ast}(892)^0\bar{K}^0$ mode at the
continuum energy $\sqrt{s}=10.52$ GeV, we can estimate the
continuum contributions at $\sqrt{s}=10.58$ GeV and 10.876 GeV under the assumption
that the continuum cross section varies as $1/s^4$. After subtracting the continuum contributions,
the net contribution to the cross sections from $\Upsilon(4S)$ and $\Upsilon(5S)$ decays is determined to be
$(-4.5\pm 3.7)$ fb and $(-0.8\pm 3.1)$ fb, respectively. Here, the errors are statistical
and systematic combined and the common systematic errors are counted once. The efficiencies and
the radiative correction factors are reevaluated assuming the events are from $\Upsilon(4S)$
or $\Upsilon(5S)$ decays, and possible interference between the continuum and resonant
amplitudes is neglected.
The total production cross sections of $\Upsilon(4S)$ and $\Upsilon(5S)$ are $(2.06\pm 0.11)$ nb
and $(0.70\pm 0.39)$ nb, calculated with the world average values of their masses and
partial widths to electron pairs~\cite{PDG}. By generating toy MC samples, assuming both the
$K^{\ast}(892)^0\bar{K}^0$ and the total production cross sections follow Gaussian distributions
(the mean values and standard deviations being set to the central values and corresponding errors
of the cross sections, respectively), we obtain the distribution of the ratio of the two cross
sections, from which the decay branching fraction upper limits
$\BR(\Upsilon(4S)\to K^{\ast}(892)^0\bar{K}^0)<2.0\times 10^{-6}$ and
$\BR(\Upsilon(5S)\to K^{\ast}(892)^0\bar{K}^0)<1.0\times 10^{-5}$ at 90\% C.L. are determined.
These results indicate that the contributions from $\Upsilon(4S)$ and $\Upsilon(5S)$ resonance
decays are insignificant.

Based on the likelihood curves of the cross section measurements,
in which the relevant systematic uncertainties are convolved, we obtain:
$$
R_{\rm VP}=\frac{\sigma_B(e^+e^-\to K^{\ast}(892)^0\bar
K^0)}{\sigma_B(e^+e^-\to K^{\ast}(892)^-K^+)}>4.3,\;\; 20.0, \;\; 5.4,
$$
and
$$
R_{\rm TP}=\frac{\sigma_B(e^+e^-\to K_2^{\ast}(1430)^0\bar
K^0)}{\sigma_B(e^+e^-\to K_2^{\ast}(1430)^-K^+)}<1.1,\;\;  0.4,\;\; 0.6,
$$
for $\sqrt{s}=10.52$ GeV, 10.58 GeV, and 10.876~GeV, respectively, at the 90\% C.L.
Assuming the cross section dependence on $s$ is $1/s^{n}$ ($n=3.83$)
from our measurement of $K^{\ast}(892)^0\bar{K}^0$ and that this assumption is applicable
to all the final states,
we obtain the weighted average of the cross sections at a luminosity-weighted energy point
of 10.61~GeV, which are $(7.86^{+0.72}_{-0.71})$ fb, $(0.54^{+0.13}_{-0.12})$ fb, $(1.36^{+0.77}_{-0.69})$ fb
and $(7.81^{+0.96}_{-0.93})$ fb
for $K^{\ast}(892)^0\bar{K}^0$,  $K^{\ast}(892)^-K^+$, $K_2^{\ast}(1430)^0\bar{K}^0$ and
$K_2^{\ast}(1430)^-K^+$, respectively.
For $K_2^{\ast}(1430)^0\bar{K}^0$ and $K_2^{\ast}(1430)^-K^+$, based on the above
weighted average of the cross sections at $\sqrt{s}=10.61$~GeV and the assumption
of the cross section dependence on $s$, we obtain
$(3.8^{+2.1}_{-1.9})$ pb and $(21.6^{+2.7}_{-2.6})$ pb  at $\sqrt{s}=3.77$~GeV.
The uncertainties are the sum in quadrature  of the statistical
and systematic uncertainties.
We obtain the averaged ratios as $\bar{R}_{\rm VP}>10.9$ and $\bar{R}_{\rm TP}<0.3$ at the 90\% C.L.
Here, for the calculated ratios, the common systematic uncertainties cancel.


For $K^{\ast}(892){\bar K}$, the ratio of the cross sections
of $K^{\ast}(892)^0\bar K^0$ and $K^{\ast}(892)^-K^+$ at $\sqrt{s}=10.58$~GeV is
much larger than the predictions from exact or broken SU(3) symmetry models. Conversely, for
$K_2^{\ast}(1430) {\bar K}$, the ratio of the cross sections of
$K_2^{\ast}(1430)^0\bar K^0$ and $K_2^{\ast}(1430)^-K^+$ is much smaller than the
prediction from the SU(3) symmetry or with the SU(3) symmetry
breaking effects considered.

In a naive quark model developed to explain the transition-rate difference between $K_2^{\ast}(1430)^0\to K^0 \gamma$ and
$K_2^{\ast}(1430)^+\to K^+ \gamma$~\cite{carl}, one obtains $R_{\rm
TP}\ll 1$ by assuming the model can be extended to a time-like
virtual-photon case; this extrapolation is justified since the same model predicted the ratio of
$\frac{\Gamma(K_2^{\ast}(1430)^0\to K^0 \gamma)}
{\Gamma(K_2^{\ast}(1430)^+\to K^+ \gamma)}=0.054$,
in rough agreement with the experimental measurement~\cite{carl}.
In the same model, however, the radiative transitions between $K^\ast(892)$ and $K$ were
also calculated, and a ratio $\frac{\Gamma(K^{\ast}(892)^0\to K^0 \gamma)}
{\Gamma(K^{\ast}(892)^+ \to K^+ \gamma)}=1.7$ was obtained, which is very
different from the measurements of $R_{\rm VP}$ from both this and
CLEO~\cite{cleo-kstrk} experiments.

In summary, we have measured for the first time the  cross sections for the reactions
$\EE \to \omega \pi^0$, $K^{\ast}(892) \bar{K}$, and $K_2^{\ast}(1430)
\bar{K}$  at CM energies between 10
and 11~GeV. The results are summarized in Table~\ref{summary}.
Significant signals of $\omega\piz$, $K^{\ast}(892)^0\bar{K}^0$, and $K_2^{\ast}(1430)^-K^+$ are
observed, while no significant excess for
$K^{\ast}(892)^-K^+$ and $K_2^{\ast}(1430)^0 \bar{K}^0$ is found.
The ratios $R_{\rm VP}$ and $R_{\rm TP}$ at the 90\% C.L. are given.


We thank the KEKB group for excellent operation of the
accelerator; the KEK cryogenics group for efficient solenoid
operations; and the KEK computer group, the NII, and
PNNL/EMSL for valuable computing and SINET4 network support.
We acknowledge support from MEXT, JSPS and Nagoya's TLPRC (Japan);
ARC and DIISR (Australia); FWF (Austria); NSFC (China); MSMT (Czechia);
CZF, DFG, and VS (Germany);
DST (India); INFN (Italy); MEST, NRF, GSDC of KISTI, and WCU (Korea);
MNiSW and NCN (Poland); MES and RFAAE (Russia); ARRS (Slovenia);
IKERBASQUE and UPV/EHU (Spain);
SNSF (Switzerland); NSC and MOE (Taiwan); and DOE and NSF (USA).



\begin{thebibliography}{**}
\bibitem{charged} Charge-conjugate decays are implicitly assumed
throughout the paper.
\bibitem{ourowncalculation} This is our own calculation based on the values of the cross sections from
Ref.~\cite{cleo-kstrk}.
\bibitem{ynso} With this energy dependence, we can calculate the branching fraction
of $\Upsilon(1S)\to \omega \pi^0$ according to $\frac{\BR(\Upsilon(1S)\to \omega\piz)}{\BR(\Upsilon(1S)\to \MM)}
=\frac{\sigma_B(\EE\to \omega\piz)}{\sigma_B(\EE\to \MM)}$ assuming
$\Upsilon(1S)$ decays into $\omega\piz$ via one-photon annihilation only.
Here the cross sections are at $\sqrt{s}=m_{\Upsilon(1S)}$, with $m_{\Upsilon(1S)}$ being
the nominal mass of the $\Upsilon(1S)$. We obtain $\BR(\Upsilon(1S)\to \omega\piz)=
(3.56^{+0.83}_{-0.78})\times 10^{-7}$, which is one order of magnitude lower than the
upper limit determined in Ref.~\cite{shen}
 and about two orders of magnitude lower than the prediction in Ref.~\cite{gerard}.
\end{thebibliography}
\end{document}